\documentclass[apjl]{emulateapj}
\usepackage{epsf}

\def\mbf#1{\mbox{\boldmath ${#1}$}}
\def\Alfven{Alfv\'{e}n~}
\def\Alfvenic{Alfv\'{e}nic~}

\slugcomment{Submitted to ApJL}

\shorttitle{Solar Wind from Photosphere to 0.3AU}
\shortauthors{T. K. Suzuki \& S. Inutsuka}

\begin{document}

\title{Making the corona and the fast solar wind: a self-consistent 
simulation for the low-frequency \Alfven waves from photosphere to 0.3AU}

\author{Takeru K. Suzuki \& Shu-ichiro Inutsuka}
\affil{Department of Physics, Kyoto University, Kitashirakawa, 
Kyoto, 606-8502, Japan}
\altaffiltext{1}{JSPS Research Fellow}

\email{stakeru@scphys.kyoto-u.ac.jp}
\email{inutsuka@tap.scphys.kyoto-u.ac.jp}

\begin{abstract}
We show that the coronal heating and the fast solar wind acceleration in the 
coronal holes are natural consequence of the footpoint fluctuations of the 
magnetic fields at the photosphere, by performing one-dimensional 
magnetohydrodynamical simulation with radiative cooling and thermal 
conduction.  
We initially set up a static open flux tube with temperature  
$10^4$K rooted at the photosphere. 
We impose transverse photospheric motions corresponding to the granulations 
with velocity $\langle dv_{\perp}\rangle = 0.7$km/s and period between 
20 seconds and 30 minutes, which generate outgoing \Alfven waves.  We  
self-consistently treat these waves and the plasma heating.  
After attenuation in the chromosphere by $\simeq 85$\% of the initial 
energy flux, the outgoing \Alfven waves enter the corona and contribute 
to the heating and acceleration of the plasma mainly by the nonlinear 
generation of the compressive waves and shocks. 
Our result clearly shows that the initial cool and static 
atmosphere is naturally heated up to $10^6$K and accelerated to 
$\simeq 800$km/s. \\
The mpeg movie for fig.1 is available at 
\begin{verbatim}
http://www-tap.scphys.kyoto-u.ac.jp/~stakeru/research/suzuki_200506.mpg
\end{verbatim}
\end{abstract}

\section{Introduction}
A certain portion of the solar surface is covered by the coronal holes.  
Their characteristic features are that the open magnetic fields 
are emanating into the interplanetary space 
and the hot plasma is streaming out. 
The high speed solar winds (800km/s at $\simeq 1$AU) are accelerated from 
the polar coronal holes which stably exists except during the solar 
maximum phase \citep{phi95}
The source of the energy for heating and accelerating the plasma is 
believed to be in the surface convection; this energy is lifted up 
through the magnetic fields and its dissipation results in the plasma heating. 
In this sense, the problem of the coronal heating and the solar wind 
acceleration is to solve how the solar atmosphere reacts to the footpoint 
motions of the magnetic fields. 

In the coronal holes the \Alfven waves can play an important role 
because it can travel a long distance to both heat the corona and  
accelerate the solar wind. 
The \Alfven waves are excited by steady transverse motions of the field 
lines at the photosphere (e.g. Cranmer \& van 
Ballegooijen 2005), while they can also be produced by reconnection 
events above the photosphere \citep{am97}. 
Although the latter process might be responsible for the generation of the 
high-frequency (up to $10^4$Hz) ioncyclotron wave highlighted 
for the heating of the heavy ions \citep{kol98}, 
it is difficult to heat the protons \citep{cra00}. 
Because we focus on the heating of the major component of the plasma, we study 
the effect of the low-frequency ($\lesssim 0.1$Hz) \Alfven waves generated 
by the steady footpoint fluctuations.

Although the \Alfven waves have been intensively investigated, 
previous studies have a couple of limitations. 
Most of the calculations consider the waves from the fixed 
`coronal base' of which density is $\sim 9$ orders of magnitudes lower than 
that of the photosphere (e.g. Oughton et al.2001; Ofman 2004).  
However, the corona dynamically interact with the chromosphere 
by chromospheric evaporation \citep{ham82} and wave transmission \citep{ks99} 
so that the coronal base varies time-dependently. 
Some calculations including the chromosphere even adopt an ad hoc 
heating function (e.g. Lie-Svendsen 2002).    
In contrast, 
we self-consistently treat the transfer of the mass/momentum/energy by solving 
the wave propagation from the 
photosphere to the interplanetary region. 
Our aim is to answer the problem of the heating and acceleration in the 
coronal holes by the {\em forward} approach \citep{gn05}.

\begin{figure}
\figurenum{1} 
\epsscale{1.} 
\plotone{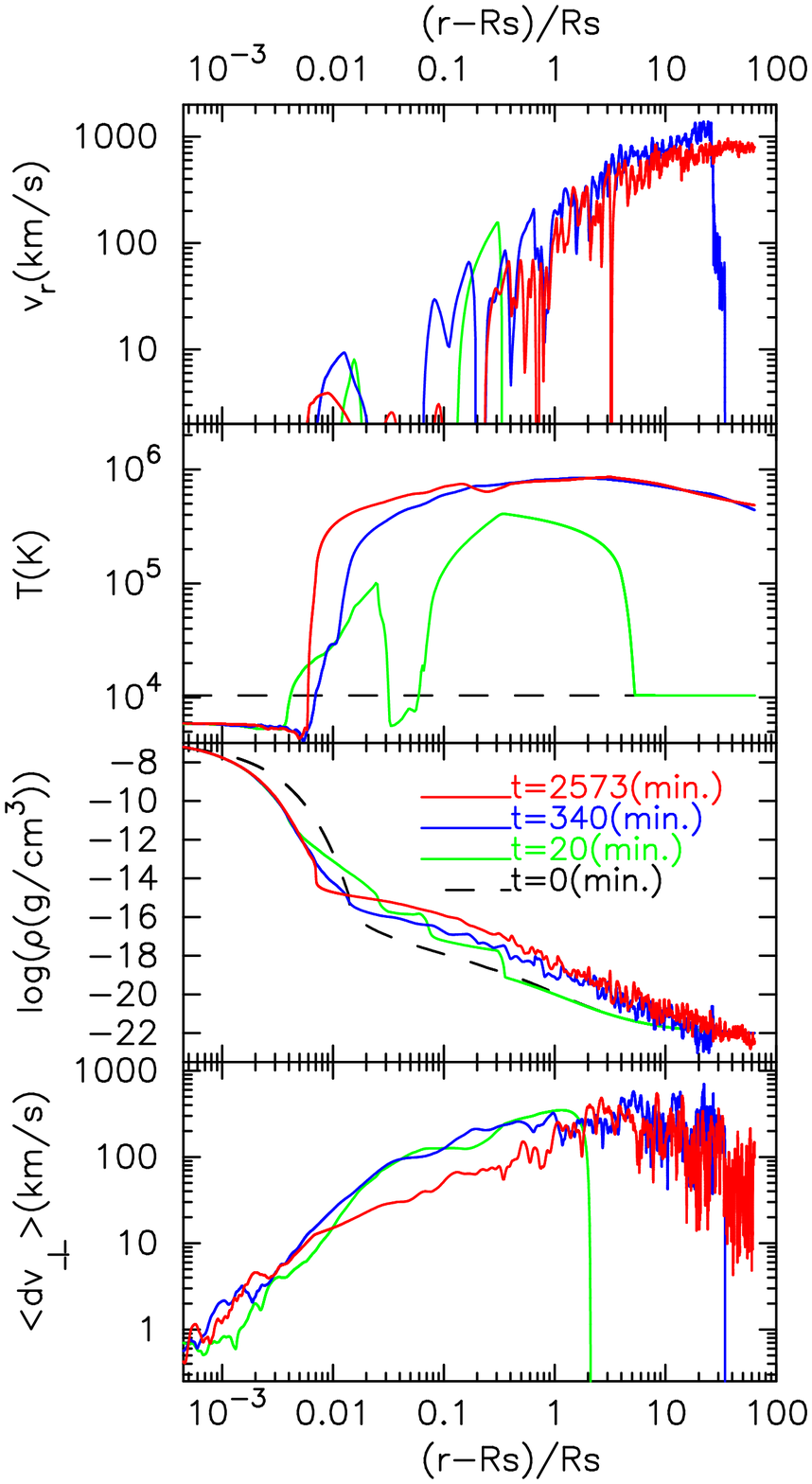} 
\caption
{Time evolution of the atmosphere.  From top to bottom, 
outflow speed, $v_r$(km/s), temperature, $T$(K),  
density, $\rho$ (g cm$^{-3}$), and rms amplitude of transverse velocity, 
$\langle dv_{\perp}\rangle$(km/s) are plotted. Dashed, 
green, blue, and red solid 
lines are results at 
$t=0$, 20, 340, \& 2573 mins., respectively, whereas results 
at $t=0$ do not appear in $v_r$ and $\langle dv_{\perp} \rangle$ 
because both equal to 0. 
The results are averaged by integrating for 3 minutes 
to take into account the effects of the exposure time for comparison with 
observations.
}
\end{figure}

\section{Method}

We consider one-dimensional (1-D) super-radially open flux tube, 
measured by heliocentric distance, $r$. 
The computation domain is from the photosphere ($r=1R_{\rm S}$) with  
density, $\rho = 10^{-7}$g cm$^{-3}$, to $65R_{\rm S}$ (0.3AU), 
where $R_{\rm S}$ is solar radius.  
Radial field strength, $B_r$, 
is given by conservation of magnetic flux as 
\begin{equation}
B_r r^2 f(r) = {\rm const.} ,
\end{equation}
where we adopt the same function as in \citet{kh76} for superradial 
expansion, $f(r)$, but consider two-step expansions: 
the flux tube initially expands by a factor of 30 at $r \simeq 
1.01R_{\rm S}$ corresponding to `funnel' structure \citep{tu05}, 
and followed by 2.5 times expansion  
at $r \simeq 1.2R_{\rm S}$ due to the large scale magnetic fields.     
$B_r$ at the photosphere is set to be 161G. These give $B_r =5$G 
at $r=1.02 R_{\rm S}$ (low corona), and 
$B_r = 2.1{\rm G}(R_{\rm S}/r)^2$ in $r>1.5 R_{\rm S}$. 
We give transverse fluctuations of the field line by the granulations at the 
photosphere, which excite \Alfven waves. 
We consider the fluctuations with power spectrum, $P(\nu)\propto \nu^{-1}$, 
in frequency between $6\times 10^{-4} \le \nu \le 0.05$Hz 
(period of 20 seconds --- 30 minutes), and root mean squared (rms) average 
amplitude $\langle dv_{\perp}\rangle\simeq 0.7$km/s corresponding to 
observed velocity amplitude $\sim 1$km/s \citep{hgr78}.
At the outer boundary, outgoing condition is imposed for all the MHD 
waves (Suzuki \& Inutsuka 2005) to avoid the unphysical wave reflection.

We treat the dynamical evolution of the waves and 
the plasma by solving ideal MHD equations with the relevant physical 
processes : 
\begin{equation}
\label{eq:mass}
\frac{d\rho}{dt} + \frac{\rho}{r^2 f}\frac{\partial}{\partial r}
(r^2 f v_r ) = 0 , 
\end{equation}
$$
\rho \frac{d v_r}{dt} = -\frac{\partial p}{\partial r}  
- \frac{1}{8\pi r^2 f}\frac{\partial}{\partial r}  (r^2 f B_{\perp}^2)
$$
\begin{equation}
\label{eq:mom}
+ \frac{\rho v_{\perp}^2}{2r^2 f}\frac{\partial }{\partial r} (r^2 f)
-\rho \frac{G M_{\rm S}}{r^2}  , 
\end{equation}
\begin{equation}
\label{eq:moc1}
\rho \frac{d}{dt}(r\sqrt{f} v_{\perp}) 
= \frac{B_r}{4 \pi} \frac{\partial} {\partial r} (r \sqrt{f} B_{\perp}).
\end{equation}
$$
\rho \frac{d}{dt}(e + \frac{v^2}{2} + \frac{B^2}{8\pi\rho} 
- \frac{G M_{\odot}}{r}) + \frac{1}{r^2 f} 
\frac{\partial}{\partial r}\left[r^2 f \left\{ (p + \frac{B^2}{8\pi}) v_r  
\right. \right.
$$
\begin{equation}
\label{eq:eng}
\left. \left.
- \frac{B_r}{4\pi} (\mbf{B \cdot v})\right\}\right]
+ 
\frac{1}{r^2 f}\frac{\partial}{\partial r}(r^2 f F_{\rm c}) 
+ q_{\rm R} = 0,
\end{equation}
\begin{equation}
\label{eq:ct}
\frac{\partial B_{\perp}}{\partial t} = \frac{1}{r \sqrt{f}}
\frac{\partial}{\partial r} [r \sqrt{f} (v_{\perp} B_r - v_r B_{\perp})], 
\end{equation}
where $\rho$, $\mbf{v}$, $p$, $e$, $\mbf{B}$ are density, velocity, pressure, 
specific energy, and magnetic field strength, respectively, and subscript 
$r$ and $\perp$ denote radial and tangential components. 
$\frac{d}{dt}$ and $\frac{\partial}{\partial t}$ denote Lagrangian and 
Eulerian derivatives, respectively.  
$G$ and $M_{\rm S}$ are the 
gravitational constant and the solar mass. $F_{\rm c}$ is thermal 
conductive flux and $q_{\rm R}$ is radiative cooling 
\citep{LM90,aa89,mor04}. 
Note that the curvature effects appear 
as $r\sqrt{f}$ terms, instead of $r$ for the usual spherical coordinate.
We adopt 2nd-order MHD-Godunov-MOCCT scheme for updating the physical 
quantities (Sano \& Inutsuka 2005),  
of which an advantage is that no artificial viscosity 
is required even for strong MHD shocks. 
We fix 14000 grid points with variable sizes in a way to resolve \Alfven waves 
with $\nu=0.05$Hz by at least $10$ grids per wavelength to reduce 
the numerical damping. 

The advantages of our calculations are automatic treatments of the wave 
propagation/dissipation and the plasma heating/acceleration without an ad hoc 
heating function but with the minimal assumption 
over the broadest regions with the huge density contrast. 
We emphasize that this is one of the most self-consistent simulations  
for the solar wind acceleration at present.
The disadvantage is the 1-D MHD approximation, 
which will be discussed later. 

\begin{figure}
\figurenum{2} 
\epsscale{0.65} 
\plotone{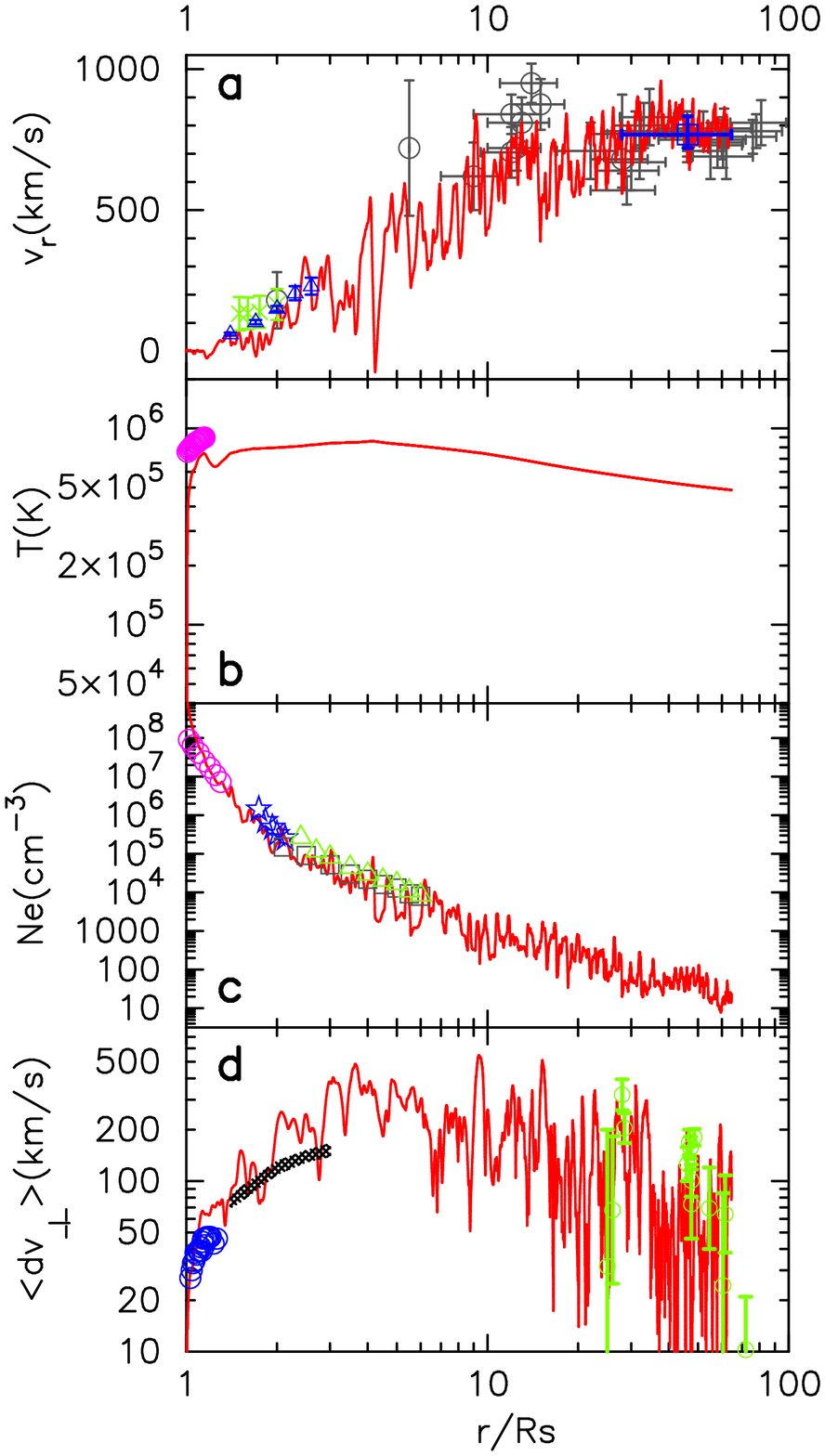} 
\caption
{Comparison of the result at $t=2573$ mins (red solid lines) 
with observations summarized below. 
Quantities in four panels are the same as in Figure 1 except 
third panel showing electron density, $N_e$(cm$^{-3}$), instead of 
$\rho$(g cm$^{-3}$). Scales 
in both horizontal and vertical axises are changed from fig.1.  
{\bf a}: 
Vertical error bars with green x's \citep{tpr03} and blue triangles 
\citep{zan02} are proton outflow speed in polar regions by SoHO. 
Blue square with errors is velocity by IPS measurements 
averaged in 0.13 - 0.3AU of high-latitude regions\citep{koj04}. 
Crossed bars with and without circles are measurements by 
VLBI and IPS(EISCAT) \citep{gra96}.  
Vertical error bars with circles are data based on observation 
by SPARTAN 201-01 \citep{hab95}. 
{\bf b}: 
Pink circles are electron temperature by CDS/SoHO \citep{fdb99}. 
{\bf c}: Pink circles and [blue] stars are observations by SUMER/SoHO 
\citep{wil98} and by CDS/SoHO \citep{tpr03}, 
respectively. [Green] triangles \citep{tpr03} 
and squares \citep{lql97} 
are observations by LASCO/SoHO. 
{\bf d}: Blue circles are non-thermal broadening inferred from 
SUMER/SoHO measurements \citep{ban98}. 
Cross hatched region 
is empirically derived non-thermal broadening based on 
UVCS/SoHO observation \citep{ess99}. 
Green error bars with circles are transverse velocity 
fluctuations derived from IPS measurements by EISCAT \citep{can02}.         
}
\end{figure}
   
\section{Results and Discussions}
We start the calculation from the static and cool atmosphere with temperature, 
$T=10^4$K; we do not set up the corona and the solar wind at the beginning.  
Figure 1 (on-line movie available) shows how the coronal heating and the 
solar wind acceleration are realized by the low-frequency \Alfven 
waves. 
We plot $v_r$(km/s), $T$(K), $\rho$(g/cm$^{3}$), 
and $\langle dv_{\perp}\rangle$(km/s)  
averaged from $v_{\perp}$ as a function of $(r-R_{\rm S})/R_{\rm S}$ 
at different time, $t=0, 20, 340$ \& 2573 minutes. 
As time goes on, the atmosphere is heated and accelerated 
effectively by the dissipation of the \Alfven waves. 
Temperature rises rapidly in the inner region even at $t=20$ minutes,    
and the outer region is eventually heated up by both outward 
thermal conduction and wave dissipation. 
Once the plasma is heated up to the coronal 
temperature, mass is supplied to the corona mainly by the chromospheric 
evaporation due to the downward thermal conduction. 
This is seen in temperature structure as an inward shift of the transition 
region, which is finally located 
around $r-R_{\rm S} =6 \times 10^{-3} R_{\rm S}$ ($\simeq$4000 km).  
As a result, the coronal density increases by two orders of magnitude. 
While the wind velocity exceeds 
1000km/s at $t=340$ minutes on account of the 
initial low density, it gradually settles down to $< 1000$ km/s  
as the density increases.   
$\langle dv_{\perp}\rangle$ also settles down to the reasonable value at the 
final stage.

We have found that the plasma is steadily heated up to $10^6$K in the 
corona and flows out as a transonic wind with $v_r \simeq 800$km/s at 
the outer boundary ($=$0.3AU) when the quasi 
steady-state behaviors are achieved after $t\gtrsim 1800$ minutes.  
This is the first numerical simulation which directly shows that 
the heated plasma actually flows out as the transonic wind, 
from the initially static and cool atmosphere, by the effects of 
the \Alfven waves. 
The sonic point where $v_r$ exceeds the local sound speed, $c_{\rm s}=
\sqrt{5p/3\rho}$, 
is located at $r\simeq 2.5R_{\rm S}$ and the \Alfven point 
for the \Alfven speed, $B_r/\sqrt{4\pi \rho}$, 
is at $r\simeq 24R_{\rm S}$. 
The obtained proton flux at 0.3AU is  
$N_p v \simeq (2 \pm 0.5) \times 10^9$(cm$^{-2}$s$^{-1}$), corresponding to 
$N_p v \simeq (1.8 \pm 0.5) \times 10^8$(cm$^{-2}$s$^{-1}$) at 1AU 
for $N_p v \propto r^{-2}$, which is consistent with the 
observed high-speed stream around the earth \citep{apr01}.

In fig.2 we compare the result at $t=2573$ minutes with 
recent observation in the high-speed solar winds from the polar regions 
by Solar \& Heliospheric Observatory (SoHO) 
\citep{zan02,tpr03,fdb99,lql97,wil98,ban98,ess99} 
and Interplanetary Scintillation (IPS) measurements \citep{gra96,koj04,can02}. 
The figure shows that our {\em forward}-approach 
simulation naturally form the corona 
and the high-speed solar wind which are observed by the dissipation of the 
low-frequency \Alfven waves.  
Although the observed wind speed around $r \simeq 10R_{\rm S}$ \citep{gra96} 
appears to exceed our result, it might reflect the wave phenomena 
in the solar wind rather than the outflow \citep{hc05}. 

\begin{figure}
\figurenum{3} 
\epsscale{1.} 
\plotone{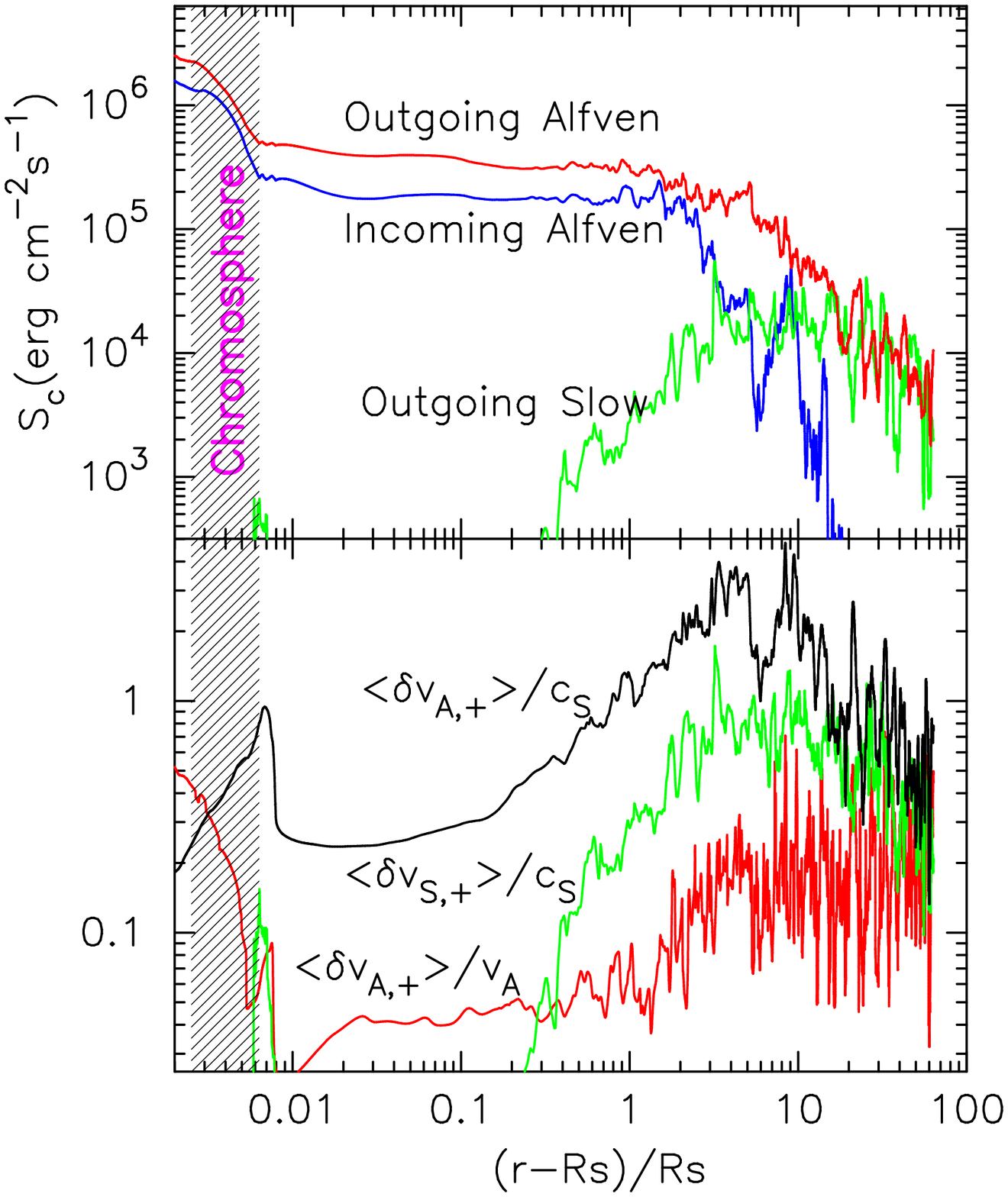} 
\caption
{{\it Top} : 
$S_c$ (see text) of outgoing \Alfven mode (red), 
incoming \Alfven mode (blue), and outgoing MHD slow 
mode (green) at 
$t=2573$ mins.  
Hatched region indicates the chromosphere and low 
transition region with $T<4\times 10^4$K.
{\it Bottom} : Normalized wave amplitude, $\langle \delta v_{\rm A,+}
\rangle /c_{\rm S}$ (solid), $\langle \delta v_{\rm A,+}
\rangle /v_{\rm A}$ (dashed [red solid]), and $\langle \delta v_{\rm S,+}
\rangle /c_{\rm S}$ (dotted [green solid]). 
}
\end{figure}

Figure 3 presents the dissipation of the waves. 
The top panel plots the following quantities, 
\begin{equation}
S_c 
=\rho \delta v^2 \frac{(v_r + v_{\rm ph})^2}
{v_{\rm ph}} \frac{r^2 f(r)}{r_c^2 f(r_c)}, 
\end{equation} 
for outgoing \Alfven, incoming \Alfven, and outgoing slow MHD (sound) waves, 
where $\delta v$ and $v_{\rm ph}$ are amplitude and phase speed of each 
wave mode, and $S_c$'s are normalized at $r_c=1.02R_{\rm s}$. 
$S_c$ is an adiabatic constant derived from wave action 
\citep{jaq77}, and corresponds to the energy flux in static media. 
$S_c$ is conserved in the expanding atmosphere if the wave does 
not dissipate, while it is not the case for the energy flux. 
For the incoming \Alfven wave in Figure 3, we plot the opposite sign of 
$S_c$ so that it becomes positive in the sub-\Alfvenic region. 
The outgoing and incoming \Alfven waves are decomposed 
by correlation between $v_{\perp}$ and $B_{\perp}$. 
Extraction of the slow wave is also from fluctuating components of $v_r$ and 
$\rho$. 
The bottom panel plots $\langle \delta v_{\rm A,+}\rangle/v_{\rm A}$, 
$\langle \delta v_{\rm A,+}\rangle/c_{\rm S}$, and 
$\langle \delta v_{\rm S,+}\rangle/c_{\rm S}$, where 
$\langle \delta v_{\rm A,+}\rangle$ and $\langle \delta v_{\rm S,+}\rangle$
are amplitudes of the outgoing \Alfven and slow modes.
All the quantities in Figure 3 are time-averaged over 30 minutes to 
smooth out the variation due to the phase. 

The top panel shows that the outgoing \Alfven waves dissipate quite 
effectively; 
$S_c$ becomes only $\sim 10^{-3}$ of the initial value 
at the outer boundary. 
First, a sizable amount is reflected back downward 
below the coronal base ($r-R_{\rm S} < 0.01 
R_{\rm S} $), which is clearly 
illustrated as the incoming \Alfven wave following 
the outgoing component with slightly smaller level.  
This is because the wave shape is considerably 
deformed owing to the steep density gradient; 
a typical variation scale ($< 10^5$km) 
of the \Alfven speed becomes comparable to or even shorter than 
the wavelength ($=10^4 - 10^6$km). 
Although the energy flux, $\simeq 5\times 10^{5}$erg cm$^{-2}$s$^{-1}$, 
of the outgoing \Alfven waves ($S_c$ in the static region 
is equivalent with the energy flux) which penetrates into the corona is 
only $\simeq 15$\% of the input value,  
it satisfies the requirement for the energy budget in the coronal holes 
\citep{wn77}. 

Second, slow MHD waves are generated in the corona 
(see Sakurai 2002 for observation) as shown in the figure. 
The amplitude of the \Alfven waves is amplified through the upward 
propagation. The bottom panel shows that $\langle \delta v_{\rm A,+}\rangle 
/c_{\rm S}\gtrsim1$ in $r\gtrsim 2R_{\rm S}$, which means that the wave 
pressure ($B_{\perp}^2/8\pi$) exceeds the gas pressure (eq.[\ref{eq:mom}])  
in spite of the weak nonlinearity, $\langle \delta v_{\rm A,+}\rangle/v_{\rm A}
\simeq 0.1$.  
This excites longitudinal slow waves \citep{ks99}, which eventually become 
nonlinear $\langle \delta v_{\rm S,+}\rangle/c_{\rm S}\simeq 1$ at 
$r\simeq 3R_{\rm S}$ to lead to the shock dissipation \citep{suz02}. 
The coronal heating and wind acceleration are thus achieved 
by transferring the energy and momentum flux of the outgoing \Alfven waves. 
Fast MHD shocks by the direct steepening of the linearly polarized components 
of the \Alfven waves also contribute to the heating \citep{hol82,suz04}, 
though they are less efficient because of the weak nonlinearity.  
The incoming \Alfven waves are generated in the corona by the reflection of 
the outgoing ones by the density fluctuations due to the slow waves. 
The reflected waves further play a role in the wave dissipation 
by nonlinear wave-wave interaction.
   
A key mechanism in the heating and acceleration of the plasma is the 
generation of the slow MHD waves. 
Thus, one of the predictions from our simulation is the existence of the 
longitudinal fluctuations in the solar wind plasma.  
This is directly testable by in situ measurements of future missions, 
Solar Orbiter and Solar Probe, 
which will approach to $\sim$45 and 4 $R_{\rm S}$, respectively, 
corresponding to the inside of our computation domain.  

We have shown by the self-consistent simulation that the dissipation of the 
low-frequency \Alfven waves through the generation of the 
compressive waves and shocks is one of the solutions for the heating and 
acceleration of the plasma in the coronal holes.    
However, the validity of the 1-D MHD approximation we have adopted 
needs to be examined. 
Generally, the shock dissipation tends to be overestimated in the 1-D 
simulation because the shocks cannot be diluted by the geometrical expansion.  
On the other hand, there are other dissipation mechanisms due to the 
multidimensionality \citep{ofm04}, such as turbulent cascade into the 
transverse direction \citep{oug01} and phase mixing \citep{hp83}. 
Therefore, the waves might also be dissipated by paths different 
from the shocks in the real situations.
The self-consistent simulations including these various processes 
remain to be done to give the final conclusion.

We thank the referee for helpful comments and 
Drs. K. Shibata and T. Sano for many fruitful discussions. 
This work is supported in part by a Grant-in-Aid for the 21st Century COE 
``Center for Diversity and Universality in Physics'' from the MEXT of Japan.
T.K.S. is supported by the JSPS Research Fellowship for Young
Scientists, grant 4607.
SI is supported by the Grant-in-Aid  (15740118, 16077202)
from the MEXT of Japan.


\begin{thebibliography}

\bibitem[Anderson \& Athay (1989)]{aa89}
Anderson, C. S. \& Athay, R. G. 
1989, \apj, 336, 1089

\bibitem[Aschwanden, Poland, \& Rabin (2001)]{apr01}
Aschwanden, M. J., Poland, A. I., \& Rabin, D. M. 
2001, \araa, 39, 175

\bibitem[Axford \& Mckenzie (1997)]{am97}
Axford, W. I. \& McKenzie, J. F. 1997, The solar wind.  
{\it Cosmic Winds and the Heliosphere}, (Eds.) Jokipii, J. R., Sonnet, C. P., 
and Giampapa, M. S., University of Arizona Press  

\bibitem[Banerjee et al.(1998)]{ban98}
Banerjee, D., Teriaca, L., Doyle, J. G., \&  Wilhelm, K. 
1998, \aap, 339, 208

\bibitem[Canals et al.(2002)]{can02}
Canals, A., Breesn, A. R., Ofman, L., Moran, P. J., \& Fallows, R. A. 
2002, Ann. Geophys. 20, 1265

\bibitem[Cranmer (2000)]{cra00}
Cranmer, S. R. 
2000, \apj, 532, 1197


\bibitem[Cranmer \& van Ballegooijen (2005)]{cb05}
Cranmer, S. R. \& van Ballegooijen, A. A. 2005, \apjs, 156, 265


\bibitem[Esser et al.(1999)]{ess99}
Esser, R., Fineschi, S., Dobrzycka, D., Habbal, S. R., Edgar, R. J., 
Raymond, J. C., \& Kohl, J. L. 
1999, \apjl, 510, L63

\bibitem[Fludra, Del Zanna, \& Bromage(1999)]{fdb99}
Fludra, A., Del Zanna, G. \& Bromage, B. J. I. 
1999, \ssr, 87, 185

\bibitem[Grall et al.(1996)]{gra96}
Grall, R. R. et al., Coles, W. A., Klinglesmith, M. T., Breen, A. R., 
Williams, P. J. S., Markkanen, J., \& Esser, R. 
1996, \nat, 379, 429

\bibitem[Gudiksen \& Nordlund (2005)]{gn05}
Gudiksen, B. V. \& Nordlund, \AA 2005, \apj, 618, 1020

\bibitem[Habbal et al.(1995)]{hab95} 
Habbal, S. R., Esser, R., Guhathakura, M., \& Fisher, R. R. 
1995, \grl, 22, 1465

\bibitem[Hammer (1982)]{ham82}
Hammer, R. 1982, \apj, 259, 767 

\bibitem[Harmon \& Coles (2005)]{hc05}
Harmon, J., K. \& Coles, W., A. 2005, \jgr, 110, A03101 

\bibitem[Heyvaerts \& Priest(1983)]{hp83}
Heyvaerts, J. \& Priest, E. R. 
1983, \aap, 117, 220

\bibitem[Hollweg (1982)]{hol82}
Hollweg, J. V. 
1982, \apj, 254. 806

\bibitem[Holweger, Gehlsen, \& Ruland (1978)]{hgr78}
Holweger, H., Gehlsen, M., \& Ruland, F. 
1978, \aap, 70, 537

\bibitem[Jacques (1977)]{jaq77}
Jacques, S. A. 1977, \apj, 215, 942

\bibitem[Kohl et al. (1998)]{kol98}
Kohl, J. L. et al. 
1998, \apjl, 501, L127

\bibitem[Kojima et al. (2004)]{koj04}
Kojima, M., Breen, A. R., Fujiki, K., Hayashi, K., Ohmi, T., \& Tokumaru, M. 
2004, \jgr, 109, A04103

\bibitem[Kopp \& Holzer(1976)]{kh76}
Kopp, R. A. \& Holzer, T. E. 1976, \solphys, 49, 43

\bibitem[Kudoh \& Shibata (1999)]{ks99}
Kudoh, T. \& Shibata, K. 
1999, \apj, 514, 493

\bibitem[Lamy et al.(1997)]{lql97}
Lamy, P. et al. 1997, 
{\it Fifth SOHO Workshop, The Corona and Solar Wind near Minimum Activity}, 
(ed) A. Wilson (ESA-SP 404; Noordwijk:ESA), 491  

\bibitem[Landini \& Monsignori-Fossi(1990)]{LM90}
Landini, M. \& Monsignori-Fossi, B. C. 
1990, \aaps, 82, 229


\bibitem[Lie-Svendsen (2002)]{lsv02}
Lie-Svendsen, \O, Hansteen, V. H., Leer, \& E. Holzer, T. E. 2002, 
\apj, 566, 562


\bibitem[Moriyasu et al.(2004)]{mor04}
Moriyasu, S., Kudoh, T., Yokoyama, T., \& Shibata, K.  
2004, \apj, 601, L107

\bibitem[Ofman (2004)]{ofm04}
Ofman, L. 2004, \jgr, 109, A07102

\bibitem[Oughton et al. (2001)]{oug01}
Oughton, S. et al., Matthaeus, W. H., Dmitruk, P., Milano, L. J., Zank, G. P., 
\& Mullan, D. J. 2001, 
\apj, 551, 565

\bibitem[Phillips et al.(1995)]{phi95}
Phillips, J. L. et al. 1995, \grl, 22, 3301

\bibitem[Sakurai et al.(2002)]{sak02}
Sakurai, T., Ichimoto, K., Raju, K. P., \& Singh, J. \solphys, 209, 265

\bibitem[Sano \& Inutsuka(2005)]{si05}
Sano, T. \& Inutsuka, S. 2005, in preparation

\bibitem[Suzuki(2002)]{suz02}
Suzuki, T. K. 2002, \apj, 578, 598

\bibitem[Suzuki(2004)]{suz04}
Suzuki, T. K. 
2004, \mnras, 349, 1227

\bibitem[Suzuki \& Inutsuka (2005)]{szi05}
Suzuki, T. K.  \& Inutsuka, S. 2005, in preparation

\bibitem[Teriaca et al.(2003)]{tpr03} 
Teriaca, L., Poletto, G., Romoli, M., \& Biesecker, D. A. 
2003, \apj, 588, 566

\bibitem[Tu et al.(2005)]{tu05}
Tu, C.-Y., Zhou, C., Marsch, E., Xia, L.-D., Zhao, L., Wang, J.-X., 
\& Wilhelm, K. 2005, Science, 308, 519

\bibitem[Wilhelm et al.(1998)]{wil98}
Wilhelm, K., Marcsh, E., Dwivedi, B. N., Hassler, D. M., 
Lemaire, P., Gabriel, A. H., \& Huber, M. C. E. 1998, 
\apj, 500, 1023

\bibitem[Withbroe \& Noyes (1977)]{wn77}
Withbroe, G. L. \& Noyes, R. W. 
1977, \araa, 15, 363

\bibitem[Zangrilli et al.(2002)]{zan02}
Zangrilli, L., Poletto, G., Nicolosi, P., Noci, G., \& Romoli, M. 
2002, \apj, 574, 477



\end{thebibliography}
\end{document}